\documentclass[superscriptaddress,twocolumn,notitlepage,pra,longbibliography]{revtex4-2}
\usepackage[colorlinks=true,citecolor=blue,urlcolor=black]{hyperref}
\usepackage{graphicx}
\usepackage{dcolumn}
\usepackage{bm}
\usepackage{amsthm}
\usepackage{amsmath}
\usepackage{amssymb}
\usepackage{color}
\usepackage{multirow}
\usepackage{algorithm}
\usepackage{algorithmicx}
\usepackage{algpseudocode}
\usepackage{subfigure}
\usepackage{hyperref}
\usepackage{wasysym}
\usepackage{xcolor}
\usepackage{overpic}
\usepackage{color}
\usepackage{lineno}
\usepackage{diagbox}
\usepackage[normalem]{ulem}

\newtheorem{definition}{Definition}
\newtheorem{theorem}{Theorem}

\definecolor{Red}{rgb}{1,0,0}

\begin{document}

\title{An effcient variational quantum Korkin-Zolotarev algorithm for solving shortest vector problems}

\author{Xiaokai Hou}
\affiliation{China Electronic Technology Cyber Security Co., Ltd., Chengdu 610041, China}
\affiliation{National Key Laboratory of Security Communication, Institute of Southwestern Communication, Chengdu 610041, China}
\affiliation{Institute of  Fundamental and Frontier Sciences, University of Electronic Science and Technology of China, Chengdu, Sichuan, 610051, China}

\author{Guoqing Zhou}
\affiliation{National Key Laboratory of Security Communication, Institute of Southwestern Communication, Chengdu 610041, China}

\author{Shan Jin}
\affiliation{Institute of  Fundamental and Frontier Sciences, University of Electronic Science and Technology of China, Chengdu, Sichuan, 610051, China}

\author{Yang Li}
\affiliation{National Key Laboratory of Security Communication, Institute of Southwestern Communication, Chengdu 610041, China}

\author{Wei Huang}
\affiliation{National Key Laboratory of Security Communication, Institute of Southwestern Communication, Chengdu 610041, China}

\author{Ao Sun}
\affiliation{National Key Laboratory of Security Communication, Institute of Southwestern Communication, Chengdu 610041, China}

\author{Xiaoting Wang}
\email{xiaoting@uestc.edu.cn}
\affiliation{Institute of  Fundamental and Frontier Sciences, University of Electronic Science and Technology of China, Chengdu, Sichuan, 610051, China}

\author{Bingjie Xu}
\email{xbjpku@163.com}
\affiliation{China Electronic Technology Cyber Security Co., Ltd., Chengdu 610041, China}
\affiliation{National Key Laboratory of Security Communication, Institute of Southwestern Communication, Chengdu 610041, China}

\begin{abstract}
Noisy intermediate-scale quantum cryptanalysis focuses on the capability of near-term quantum devices to solve the mathematical problems underlying cryptography, and serves as a cornerstone for the design of post-quantum cryptographic algorithms. For the shortest vector problem (SVP), which is one of the computationally hard problems in lattice-based cryptography, existing near-term quantum cryptanalysis algorithms map the problem onto a fully-connected quantum Ising Hamiltonian, and obtain the solution by optimizing for the first excited state. However, as the quantum system scales with the problem size, determining the first excited state becomes intractable due to the exponentially increased complexity for large-scale SVP instances. In this paper, we propose a variational quantum Korkin-Zolotarev (VQKZ) algorithm, which significantly reduces the qubit requirement for solving the SVP.  Specifically, by transforming the original SVP into a series of subproblems on projected sublattices, the proposed VQKZ algorithm enables near-term quantum devices to solve SVP instances with lattice dimensions $61.39\%$ larger than those solvable by previous methods. Furthermore, numerical simulations demonstrate that the proposed VQKZ algorithm can significantly outperform existing methods in terms of the length of solution vectors.

\end{abstract}

\maketitle

\section{\label{sec:Introduction} Introduction}

Cryptanalysis aims to study the vulnerabilities of cryptographic algorithms for decrypting ciphertext. Traditional cryptanalysis methods include brute-force attacks~\cite{tezcan2022key, mok2019intelligence, kumar2011investigations}, statistical analysis~\cite{rizzi2009statistical, vaudenay1996experiment, xu2021applications}, theoretical attacks~\cite{matsui1993linear, langford1994differential, biham1992differential, mantin2005predicting}, side-channel attacks~\cite{standaert2010introduction, spreitzer2017systematic, randolph2020power}, and so on. Among them, theoretical attacks exploit the mathematical properties of cryptographic algorithms to design attacks, which can be used to theoretically evaluate the security of the algorithms. Meanwhile, quantum theoretical attack has become an emerging interdisciplinary field which combines cryptanalysis with quantum computing, aiming to achieve quantum speedup in cryptanalysis. It has been demonstrated that Shor's algorithm achieves an exponential speedup in the cryptanalysis of Rivest-Shamir-Adleman~(RSA) encryption and elliptic curve cryptography~(ECC), which are based on problems intractable for classical algorithms~\cite{shor1999polynomial, shor1994algorithms}. In addition, recent studies show that many quantum algorithms can solve cryptographic problems with certain levels of advantages, ranging from Grover algorithm~\cite{grassl2016applying, jaques2020implementing, fernandez2024implementing, sarah2024practical, cai2023quantum, chen2025quantum, chailloux2017efficient}, Simon algorithm~\cite{kaplan2016breaking, santoli2016using, xiang2025quantum, canale2022simon, cui2021applications, dong2019quantum}, HHL algorithm~\cite{ding2023limitations,chenyuao2025quantum, roy2024new}, to quantum random walk~\cite{drissi2023new, chailloux2021lattice, bonnetain2023finding}, Grover-meets-Simon algorithm~\cite{sun2023quantum, zhou2023breaking, zhang2022quantum, xu2023quantum, zhang2024superposition, guo2022quantum} and quantum machine learning~\cite{kim2021cryptanalysis, ajith2022frequency, kim2023quantum, dai2021quantum, hariharasitaraman2024qhopnn}. These results pose significant threats to traditional cryptographic schemes.

To resist quantum computing attacks, lattice-based cryptography, as a typical class of post-quantum cryptography, utilizes hard computational problems based on lattices to design cryptographic algorithms. These hard problems include the shortest vector problem~(SVP), the closest vector problem~(CVP), and the learning with errors~(LWE) problem. Specifically, SVP involves finding the shortest non-zero vector in a Euclidean lattice and is proven to be NP-hard in the worst case~\cite{10.1145/276698.276705, van1981another}. This problem has been widely applied in the design of encryption algorithms~\cite{hoffstein1998ntru,bos2018crystals, yang2023fine, prest2020falcon}, digital signatures~\cite{hoffstein2003ntrusign, ducas2018crystals, luc2023building}, and key exchange protocols~\cite{jun2005novel, pablos2022design, lei2013ntru}. To solve SVP, quantum approaches to enhance classical methods were proposed, including the use of Grover's algorithm~\cite{laarhoven2015finding} and quantum random walks~\cite{chailloux2021lattice, bonnetain2023finding} to achieve quantum speedup for classical sieve methods, as well as the application of quantum random-access memory (QRAM) to reduce the space complexity of classical algorithms~\cite{kirshanova2019quantum, doriguello2024practicality}. 

With the advent of the noisy intermediate-scale quantum (NISQ) era~\cite{preskill2018quantum}, NISQ algorithms have been developed to leverage near-term quantum devices for addressing computational problems in  quantum simulation~\cite{kandala2017Hardwareefficient,huang2024Adaptive, ma2025Experimental}, combinatorial optimization~\cite{li2025efficient,wu2025Resourceefficient, zou2025Multiscale}, machine learning~\cite{hou2023duplicationfree, li2024Ensemblelearning, wu2025Quantum}, among other areas. Recently, there has been growing interest in evaluating whether the near-term quantum devices can pose challenges to cryptographic problems~\cite{lv2022Using, zheng2024quantumclassical, joseph2020not, joseph2021two, zhu2022iterative, albrecht2023variational}. For SVP, existing proposals transform the SVP into the problem of searching the first excited state of a fully-connected quantum Ising model $H$, and leverage quantum adiabatic computing~\cite{joseph2020not}, quantum annealing~\cite{joseph2021two},  quantum approximate optimization algorithm (QAOA)~\cite{zhu2022iterative}, and the variational quantum eigensolver (VQE)~\cite{albrecht2023variational} to obtain the shortest vector. However, the number of qubits of $H$ increases polynomially with the problem size (lattice rank) of the SVP. Due to the volume law of entanglement entropy in quantum systems~\cite{bianchi2022volume}, the complexity of representing the first excited state grows exponentially with the system size. This makes it computationally intractable to determine the first excited state for large-scale SVP instances. 

In this paper, we propose a variational quantum Korkin-Zolotarev (VQKZ) algorithm based on VQE and rigorously prove an upper bound on the number of qubits required for its implementation. Our proposal transforms the original SVP with the lattice rank $r$ into a series of subproblems on its projected sublattices with rank $\beta$ ($\beta<r$) and can effectively generate a $\beta$-block Korkin-Zolotarev (BKZ) reduced basis to solve the SVP with only $\lfloor\frac{1}{4}(\beta^2 + 3\beta)\rfloor$ qubits, significantly reducing the dependence of the number of qubits on the lattice rank compared to existing algorithms such as the quantum Ising algorithm (QIA)~\cite{joseph2021two} and the iterative quantum optimization algorithm (IQOA)~\cite{zhu2022iterative}, which require $1.5r\log_2r+r$ qubits to solve the SVP. Theoretical analysis shows that, based on existing quantum computing devices, VQKZ can solve SVP instances that are $61.39\%$ larger in scale compared to those solvable by QIA and IQOA. Furthermore, numerical simulation results demonstrate that our VQKZ can solve typical SVP instances with shorter vectors than QIA and IQOA, requiring only $6$ qubits compared to the $26$ qubits required by these two algorithms.

\section{\label{sec:lattice_svp}Shortest vector problem and basis reduction}

Let us briefly review the concept of the SVP and the basis reduction. An $m$-dimensional lattice $\mathcal{L}$, defined as a discrete subset of $\mathbb{R}^m$, can be generated by a lattice basis consisting of linearly independent vectors $\bm{b}_1,\cdots,\bm{b}_r\in \mathbb{R}^m$, where $r$ is the rank of $\mathcal{L}$ (with $r \leq m$). Formally, the lattice can be given as
\begin{equation}
  \mathcal{L}(\bm{b}_1, \cdots, \bm{b}_r) := \left\{\sum_{i=1}^{r}v_i\bm{b}_i:v_i\in \mathbb{Z} \right\}.
\end{equation}
By arranging these basis vectors as columns of a matrix $B=[\bm{b}_1,\cdots,\bm{b}_r]\in \mathbb{R}^{m\times r}$, the lattice can equivalently be expressed as
\begin{equation}
  \mathcal{L}(\bm{B}) = \left\{\bm{B} \bm{v}: \bm{v} \in \mathbb{Z}^r \right\}.
\end{equation}

The shortest vector problem (SVP) for a lattice $\mathcal{L}(B)$ is to find a nonzero vector $\bm{v}\in\mathcal{L}(B)$ with minimal $\ell_2$-norm, i.e.,
\begin{equation}
\bm{v} = \text{argmin}_{\bm{x} \in \mathbb{Z}^r \setminus {\bm{0}}} \|B\bm{x}\|.
\end{equation}
Since a lattice admits infinitely many distinct bases, solving SVP requires constructing a high-quality basis from a given basis $B$, followed by determining the shortest vector, which is far from trivial. In fact, solving SVP has been proven to be NP-hard in the worst case~\cite{10.1145/276698.276705, van1981another}.

A high-quality basis refers to a set of vectors with short norms and pairwise near-orthogonality. Such a basis can be obtained through refining the original basis $B$ with a process called basis reduction. In most basis reduction methods, Gram-Schmidt orthogonalization (GSO) is an integral step that transforms the original basis vectors $\{\bm{b}_i\}_{i=1}^r$ into an orthogonal set $\{\bm{b}^{*}_i\}_{i=1}^r$,  defined recursively by
\begin{equation}
    \bm{b}_i^*:=\bm{b}_i-\sum_{j<i}\mu_{i,j}\bm{b}_j^* \text{~~with~} 1\leq j< i \leq r
\end{equation}
where $\mu_{i,j}=\frac{\bm{b}_i\cdot \bm{b}_j^*}{\bm{b}_j^*\cdot \bm{b}_j^*}$ and $\bm{a}\cdot \bm{b}$ denotes the inner product of vectors $\bm{a}$ and $\bm{b}$. Besides GSO, another frequently used operation is the orthogonal projection,
\begin{equation}
  \pi_i:\mathbb{R}^m \rightarrow \text{span}(\bm{b}_1,\cdots,\bm{b}_{i-1})^{\perp},
\end{equation}
explicitly defined for $1\leq i \leq r$ as
\begin{equation}
    \pi_i(\bm{x}):=\sum_{j=i}^r\frac{\bm{x}\cdot\bm{b}_j^*}{\bm{b}_j^*\cdot\bm{b}_j^*} \bm{b}_j^*.
\end{equation}
Using the orthogonal projection, GSO can be written as $\bm{b}_i^*=\pi_i(\bm{b}_i)$. 

As the output of basis reduction methods, reduced bases include the $\delta$-Lenstra–Lenstra–Lov\'asz~(LLL) reduced basis~\cite{lenstra1982factoring}, $\beta$-block Korkin-Zolotarev~(BKZ) reduced basis~\cite{schnorr1994lattice}, Hermite Korkine-Zolotarev~(HKZ) reduced basis~\cite{korkine1873formes}, etc. Among them, due to its polynomial-time computability, the $\delta$-LLL reduced basis serves as the foundation for various reduction algorithms and can be formally defined as follows.
\begin{definition}
  A lattice basis $B=[\bm{b}_1,\cdots,\bm{b}_r]\in\mathbb{R}^{m\times r}$ is $\delta$-LLL reduced if
  \begin{itemize}
  \item \textbf{Size-reduced condition}: $|\mu_{i,j}|\leq\frac{1}{2}$ for all $1\leq j < i \leq r$,
  \item \textbf{Lovasz condition}: $\delta\|\bm{b}_{i-1}^*\|^2\leq \|\bm{b}_i^* + \mu_{i,i-1}\bm{b}_{i-1}^*\|^2$ for $1<i\leq r$, 
  \end{itemize}
  where $\frac{1}{4}<\delta<1$. 
\end {definition}
Despite its efficiency, the shortest vector $\bm{b}_1$ in a $\delta$-LLL reduced basis only satisfies $\|\bm{b}_1\|^2 \leq 2^{r-1} \lambda_1^2$, remaining exponentially larger than the norm of shortest vector $\lambda_1$ on the lattice. To further tighten the bound, the $\beta$-BKZ reduced basis achieves
\begin{equation}
    \|\bm{b}_1\|^2 \leq \alpha_{\beta}^{\frac{r-1}{\beta-1}} \lambda_1^2 \quad (\beta \geq 2),
\end{equation}
where $\beta$ is a hyperparameter denoting the block size (or the rank of sublattice) and $\alpha_{\beta} := \max_{1 \leq j \leq \beta} \|\bm{b}_1\|^2/\|\bm{b}_j^*\|^2$, but requires extra cost in complexity. 

For a lattice basis $B = [\bm{b}_1, \ldots, \bm{b}_r]$, let $\mathcal{L}(\hat{B}_{[j,k]}) := \mathcal{L}\big(\pi_j(\bm{b}_j), \ldots, \pi_j(\bm{b}_k)\big)$ denote the projected sublattice obtained by projecting the basis vectors $\bm{b}_j, \ldots, \bm{b}_k$ onto the orthogonal complement of $span(\bm{b}_1, \ldots, \bm{b}_{j-1})$, where $1\leq j < k \leq r$ and $k = \min(j+\beta-1, r)$. A $\beta$-BKZ reduced basis is formally defined as follows.
\begin{definition}\label{def:BKZ_reduced}
For an integer $\beta\geq 2$, a basis $B=[\bm{b}_1, \cdots, \bm{b}_r]$ is called $\beta$-BKZ reduced if it satisfies the following conditions
\begin{itemize}
\item It is size-reduced, i.e., $|\mu_{i,j}|\leq\frac{1}{2}$ for all $1\leq j < i \leq r$,
\item For each $j=1,\cdots r-1$, it holds that
\begin{equation*}
  \| \bm{b}_j\| = \lambda_1(\mathcal{L}(\hat{B}_{[j,k]})).
\end{equation*}
\end{itemize}
\end{definition}

This definition implies that the $j$-th vector in a $\beta$-BKZ reduced basis is the shortest vector of the rank-$\beta$ projected sublattice $\mathcal{L}(\hat{B}_{[j,k]})$. To construct such a basis, the BKZ algorithm~\cite{schnorr1994lattice} iteratively invokes an SVP oracle on $\mathcal{L}(\hat{B}_{[j,k]})$. However, existing classical SVP oracles exhibit exponential complexity in the rank $\beta$. For instance, enumeration requires $2^{O(\beta^2)}$ operations and sieving requires $2^{0.2972\beta}$ operations. This leads to solving the SVP with BKZ algorithm becoming intractable as $\beta$ increases. To address this problem, we propose the VQKZ algorithm which employs VQE to construct an effective quantum SVP oracle, aiming to leverage potential quantum advantages to accelerate SVP solution.

\section{\label{sec:VQKZ}Variational Quantum Korkin-Zolotarev algorithm}

In this section, we construct a quantum SVP oracle based on the VQE and propose the VQKZ algorithm to generate the $\beta$-BKZ reduced basis for solving SVP with only $\lfloor\frac{1}{4}(\beta^2 + 3\beta)\rfloor$ qubits.

\subsection{\label{subsec:framework}VQKZ framework}

For a given lattice basis $B=[\bm{b}_1,\bm{b}_2,\cdots,\bm{b}_r]$, the VQKZ algorithm aims to transform it into a $\beta$-BKZ reduced basis with a predefined block size $\beta$, ensuring that for $1\leq j \leq r-1$, the vector $\bm{b}_j$ is the shortest vector in the projected sublattice $\mathcal{L}(\hat{B}_{[j,k]})$, and all Gram-Schmidt coefficients satisfy $|\mu_{ij}|\leq \frac{1}{2}$, according to Def. \ref{def:BKZ_reduced}. To achieve this goal, the VQKZ algorithm starts with a classical LLL reduction~\cite{lenstra1982factoring}, followed by an iterative process that invokes the quantum SVP oracle to find the shortest vector in each projected sublattice. If the obtained vector already exists in the basis, LLL reduction is applied to ensure the Gram-Schmidt coefficients satisfy the size-reduced condition; otherwise, the vector is inserted into the basis, and LLL reduction is applied to remove linear dependencies and restore linear independence.

Specifically, in the initialization of the VQKZ algorithm, we denote $j$ as the starting index of the current block (i.e., the projected sublattice), and $z$ as a counter tracking the number of basis vectors satisfying the $\beta$-BKZ reduction criteria. We then use the LLL reduction to preprocess the lattice basis $B = [\bm{b}_1, \cdots, \bm{b}_r]$, such that it satisfies $|\mu_{i,j}| \leq \frac{1}{2}$ for all $1 \leq j < i \leq r$. Subsequently, we implement an iterative process to solve SVP on blocks until $z = r - 1$. Within each iteration, we update the block starting index $j$, the block ending index $k$, and the starting index of the next block $h$. The basis vectors $[\bm{b}_j,\cdots, \bm{b}_k]$ are then projected onto the sublattice $\mathcal{L}(\hat{B}_{[j,k]})$, followed by invoking the quantum SVP oracle to obtain the shortest vector $\bm{v}^{(j)}$ in this projected sublattice. If $\bm{v}^{(j)}=[1,0,\cdots,0]^T$, indicating that $\bm{b}_j$ already represents the shortest vector in $\mathcal{L}(\hat{B}_{[j,k]})$, we increment $z$ by $1$  and apply LLL reduction to the first $h$ basis vectors to maintain size reduction. Otherwise, we insert $\sum_{i=j}^{k} v^{(j)}_i\bm{b}_i$ into the basis at position $j$, and apply LLL reduction to the first $h+1$ vectors to eliminate linear dependency. Since after the insertion and LLL reduction, the updated basis vectors up to index $j$ only satisfy size reduction without necessarily meeting the shortest vector conditions for their respective projected sublattices, we reset the counter $z$ to zero. The complete workflow of the VQKZ algorithm is formalized in Algorithm~\ref{alg:VQKZ}.

\begin{algorithm}[H]
  \caption{Variational quantum Korkin-Zolotarev algorithm}
  \label{alg:VQKZ}
  \begin{algorithmic}[1]
    \Require A basis $B=[\bm{b}_1,\bm{b}_2,\cdots,\bm{b}_r]$, block size $\beta\geq 2$, LLL reduction parameter $\delta\in(0.25,1)$;
    \Ensure $\beta$-BKZ reduced basis $[\bm{b}_1,\bm{b}_2,\cdots,\bm{b}_r]$;
    \State $z \gets 0$, $j \gets 0$;
    \State $[\bm{b}_1,\bm{b}_2,\cdots,\bm{b}_r] \gets $ LLL$(\delta, [\bm{b}_1,\bm{b}_2,\cdots,\bm{b}_r])$;
    \While{$z < r-1$}
    \State $j\gets (j\mod (r-1)) +1$;          
    \State $k\gets \min(j+\beta-1, r)$;
    \State $h\gets\min(k+1,r)$;
    \State $[\hat{\bm{b}}_j,\hat{\bm{b}}_{j+1},\cdots,\hat{\bm{b}}_k]\gets\pi_j(\bm{b}_j,\bm{b}_{j+1},\cdots,\bm{b}_k)$
    \State $\bm{v}^{(j)} \gets$ quantum SVP oracle$([\hat{\bm{b}}_j,\hat{\bm{b}}_{j+1},\cdots,\hat{\bm{b}}_k])$ 
    \If{$\bm{v}^{(j)}\neq[1,0,\cdots,0]^T$}
        \State $z\gets 0$;
        \State $[\bm{b}_1,\cdots,\bm{b}_h]\gets$ LLL$(\delta, [\bm{b}_1,\cdots,\sum_{i=j}^{k}v^{(j)}_i \bm{b}_i,$\\~~~~~~~~~~~~~~~~~~~~~~~~~~~~~~~~~~~$\bm{b}_j,\cdots,\bm{b}_h])$;
      \Else
        \State $z\gets z+1$;
        \State $[\bm{b}_1, \bm{b}_2,\cdots, \bm{b}_h]\gets$ LLL$(\delta, [\bm{b}_1, \bm{b}_2,\cdots, \bm{b}_h])$

    \EndIf
    \EndWhile
  \end{algorithmic}
\end{algorithm}

\subsection{\label{quantum_svp_oracle}Quantum SVP oracle construction}

For the VQKZ algorithm,  its performance critically depends on the effectiveness of the quantum SVP oracle. In this subsection, we utilize the variational quantum eigensolver (VQE) to construct a quantum SVP oracle for solving the SVP on projected sublattices. Additionally, since the SVP solution must be a nonzero vector, we introduce an excited-state generation method as shown in Ref.~\cite{Higgott2019variationalquantum} into the oracle design to prevent trivial solutions.

In the VQE framework,  for a given Hamiltonian $H$, VQE  aims to minimize the expectation value 
\begin{equation}
\langle H \rangle:=\langle 0|U^{\dagger}(\bm{\theta})HU(\bm{\theta})|0\rangle,
\end{equation}
thereby obtaining the ground-state energy $E_{\text{gs}}$ and the corresponding ground state $|\psi_{\text{gs}} \rangle$ through iterative optimization of the parameters $\bm{\theta}$ of a circuit ansatz $U(\bm{\theta})$. To adapt VQE for constructing an SVP oracle, we need to map the SVP on a projected sublattice $\mathcal{L} (\hat{B}_{[j,k]})$ to a target Hamiltonian $H$ which can be solved by VQE.

For $\mathcal{L}(\hat{B}_{[j,k]})$, labeling its basis $\pi_j(\bm{b}_i)$, $j\leq i\leq k$, as $\hat{\bm{b}}_i$, we can represent an arbitrary vector $\bm{v}\in\mathcal{L}(\hat{B}_{[j,k]})$ as $\bm{v}=\sum_{i=j}^k x_i \hat{\bm{b}}_i$ with $x_i\in\mathbb{Z}$. Considering the $\ell_2$ norm of $\bm{v}$, we obtain
\begin{align}\label{Eq:l2norm}
	\|\bm{v}\|^2 &= || \hat{B}_{[j,k]} \bm{x}||^2\\
			 &= \bm{x}^{\text{T}}\hat{B}^{\text{T}}_{[j,k]} \hat{B}_{[j,k]} \bm{x}\\
			 &= \sum_{l=j}^k \sum_{s=j}^k x_l x_s \hat{\bm{b}}_l^{\text{T}}\cdot \hat{\bm{b}}_s.
\end{align}
Without loss of generality, we assume the range of $x_i$ satisfying $[-2^{q-1}+1, 2^{q-1}]$ where $q\in\mathbb{N}^+$. We can replace $x_i$ by constructing an $q$-qubit operator $Q_i$ as
\begin{equation}\label{Eq:Q_i}
  Q_i = \frac{1}{2}\left( \sum_{s=1}^q 2^{s-1} Z_s^{(i)} + \mathbb{I}\right)
\end{equation}
which has an integral spectrum ranging from $-2^{q-1}+1$ to $2^{q-1}$, and further construct the target Hamiltonian as
\begin{equation}\label{Eq:H}
  H = \sum_{l=j}^k \sum_{s=j}^k \hat{\bm{b}}_l^{\text{T}}\cdot \hat{\bm{b}}_s Q_l Q_s.
\end{equation}
By this way,  we can find that $H$ is a diagonal matrix, which implies that its eigenstates is the computational basis mutually composed of the eigenstates of the operators $\{Q_i\}_{i=j}^k$.  Denoting the eigenstate of $Q_i$ as $|\hat{u}_i\rangle$ with the corresponding eigenvalue $x_i$, we can obtain the eigenstate $|u\rangle$ of $H$ as 
\begin{equation}
  |u\rangle = |\hat{u}_j\rangle\otimes |\hat{u}_{j+1}\rangle\otimes\cdots\otimes|\hat{u}_{k}\rangle,
\end{equation}
and the expectation value $\langle H\rangle$ as
\begin{align}
  \langle u|H|u\rangle &= \sum_{l=j}^k \sum_{s=j}^k   \hat{\bm{b}}_l^{\text{T}}\cdot \hat{\bm{b}}_s \langle u| Q_l Q_s |u\rangle \\
  &= \sum_{l=j}^k \sum_{s=j}^k x_l x_s \hat{\bm{b}}_l^{\text{T}}\cdot \hat{\bm{b}}_s\\
  &=\|\bm{v}\|^2.
\end{align}
We can obtain the shortest vector in $\mathcal{L}(\hat{B}_{[j,k]})$ through iteratively minimizing $\langle H\rangle$. More analysis about the total number of qubits will be shown in Sec.\ref{sec:complexity_analysis}. 

However, since the expectation value of $H$ equals the norm of $\bm{v}$, the ground state of $H$ corresponds to the zero vector, which conflicts with the requirement of SVP. For solving SVP, we need to determine the first-excited state and adopt the following loss function in VQKZ,
\begin{equation}
  L(\bm{\theta}) = \langle 0|U^{\dagger}(\bm{\theta})HU(\bm{\theta})|0\rangle+\gamma |\langle0|U^{\dagger}(\bm{\theta})| \psi_{\text{gs}}\rangle|^2,
\end{equation}
where $|\psi_{\text{gs}}\rangle$ indicats the ground state of $H$ and $\gamma$ is a coefficient satisfying $\gamma\geq E_1-E_0$ with the ground energy $E_0$ and the first-excited energy $E_1$. 

For $|\psi_{\text{gs}}\rangle$, the zero vector implies that $x_i$ in $\bm{v}$ is $0$ corresponding to the eigenstate $|0\cdots01\rangle$ of an $q$-qubit operator $Q_i$. We can express $|\psi_{\text{gs}}\rangle$ in 
\begin{equation}
  |\psi_{\text{gs}}\rangle = \underbrace{\overbrace{|0\cdots01\rangle}^{Q_j}\otimes\cdots \otimes\overbrace{|0\cdots01\rangle}^{Q_k}}_{\beta q \text{ qubits}},
\end{equation}
and generate it with a circuit $U_{\text{gs}}$ consisting of $\beta$ Pauli-$X$ gates. Furthermore, we can obtain the overlap as 
\begin{align}
  |\langle0|U^{\dagger}(\bm{\theta})| \psi_{\text{gs}}\rangle|^2 & = \langle0|U^{\dagger}(\bm{\theta})| \psi_{\text{gs}}\rangle\langle \psi_{\text{gs}}|U(\bm{\theta})|0\rangle\\
  &= \langle0|U^{\dagger}(\bm{\theta})U_{\text{gs}}|0\rangle\langle 0|U^{\dagger}_{\text{gs}}|U(\bm{\theta})|0\rangle\\
  &=\langle0|U^{\dagger}(\bm{\theta})U_{\text{gs}}|0\rangle\langle 0|U_{\text{gs}}|U(\bm{\theta})|0\rangle,
\end{align}
which is the expectation value of $|0\rangle\langle0|$ with the state $|\psi\rangle = U_{\text{gs}}U(\bm{\theta})|0\rangle$. For the coefficient $\gamma$, according to the definition of $\ell_2$ norm, the ground energy $E_0$ is zero. Meanwhile, since the eigenstates of $H$ are all computational basis, we can randomly choose one computational basis except $|\psi_{\text{gs}}\rangle$ and further measure the expectation value of $H$, which can be taken as $\gamma$. By this way, we can construct an quantum SVP oracle in VQKZ for effectively solving SVP.

\subsection{SVP ansatz design and classical post-processing to enhance the SVP oracle}

\begin{figure}[H]
  \centering
  \includegraphics[width=1\linewidth]{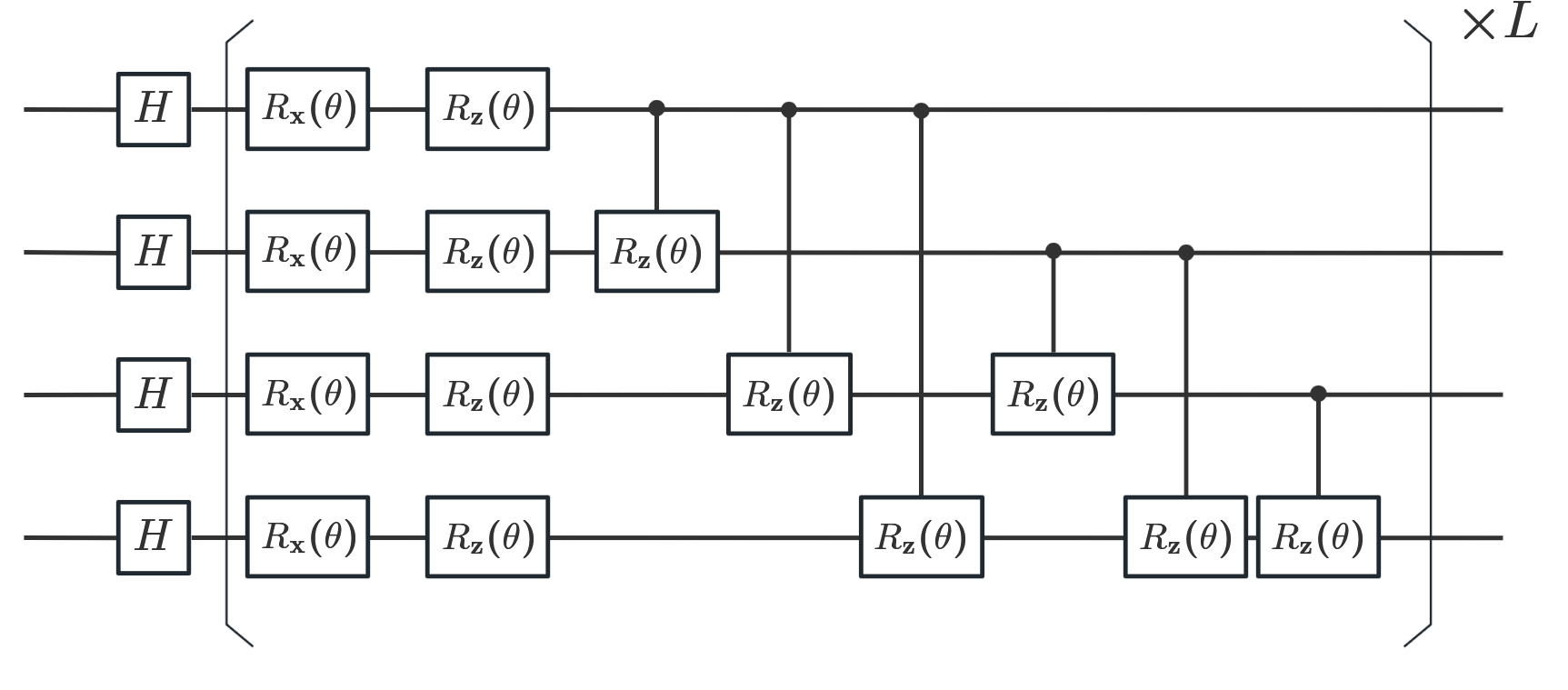}
  \caption{An QAOA-inspired SVP ansatz with $4$ qubits and $L$ layers. Following a single-layer circuit composed of $H$ gates, there are $L$ layers consisting of $R_x(\theta)$, $R_z(\theta)$ and $CR_z(\theta)$.}
  \label{oracle_ansatz}
\end{figure}

As a VQE-based quantum SVP oracle, its effectiveness depends on whether the circuit ansatz can efficiently generate the first-excited state of $H$. Quantum adiabatic evolution, as a method that maintains a quantum system in its ground state by gradually adjusting the Hamiltonian, has been demonstrated to enhance the transition probability from the ground state to the first excited state when the evolution time is shortened~\cite{joseph2020not}. In the meanwhile, the quantum approximate optimization algorithm (QAOA), which simulates quantum adiabatic evolution through alternating applications of a simple Hamiltonian $H_m$ and the problem Hamiltonian $H$ can effectively reduce computational time. However, in solving SVP, the inherent complexity of the Hamiltonian $H$ poses a significant challenge in generating its corresponding unitary evolution. To address this, we design a QAOA-inspired SVP ansatz in the SVP oracle as shown in Fig.~\ref{oracle_ansatz}.

Considering the implement of a $L$-layer QAOA, the output state can be given as
\begin{equation}
  |\psi(\bm{\gamma}, \bm{\alpha})\rangle = e^{-i\alpha_L H}e^{-i\gamma_L H_m}\cdots e^{-i\alpha_1 H}e^{-i\gamma_1 H_m}|\psi\rangle,
\end{equation}
where $H$ and $H_m$ indicate the problem Hamiltonian and the mix Hamiltonian, $\bm{\gamma}$ and $\bm{\alpha}$ indicate the trainable parameters. Typically, $H_m$ is set to $H_m=\sum_i X_i$, and $|\psi\rangle$ is prepared as $|+\rangle$. To design the SVP ansatz, we first apply a Hadamard gate to each qubit, initializing the system into $|+\rangle$. Subsequently, we parametrize the operators $e^{-i\gamma H_m}$ and $e^{-i\alpha H}$ to construct a quantum circuit with $L$ layers. In each layer, we apply a $R_x(\theta)$ gate to each qubit for simulating $e^{-i\gamma H_m}$, followed by an $R_z(\theta)$ gate to simulate the unitary generated by $Q_i^2$ terms in $H$. Finally, we implement controlled-$Rz$ ($CR_z(\theta)$) gates to simulate the unitary generated by $Q_iQ_j$ coupling terms in $H$. More comparisons between the SVP ansatz and the hardware-effcient ansatz widely used in existing works for solving the SVP problems are shown in Appendix~\ref{appendix:comparison}.

\begin{figure*}[htp]
    \setlength{\belowcaptionskip}{-2mm}
    \centering
    
    \includegraphics[width=0.6\linewidth]{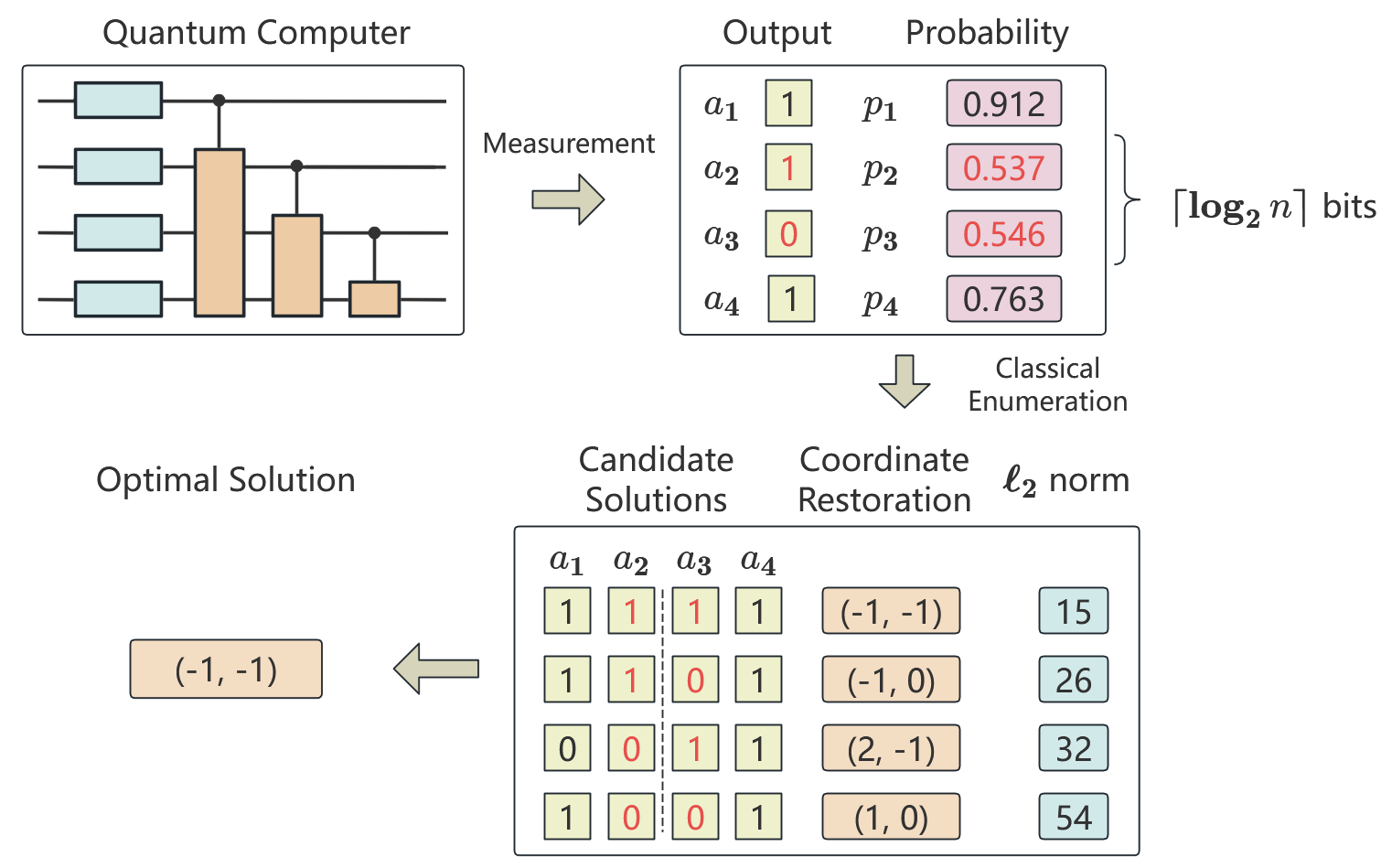}
    
    \caption{An illustraion of the post-processing scheme with $n=4$ qubits. After measurement, $2$ bits with probabilities closed to $0.5$ are chosen for generating the candidate solutions by the classical enumeration. The optimal solution is generated by comparing the $\ell_2$ norm of the candidate solutions.}
    \label{post_process}
\end{figure*}

To further enhance the SVP oracle, we design a classical post-processing scheme for the SVP oracle, as shown in Fig.~\ref{post_process}. After optimizing the circuit parameters, when performing measurements on the quantum system, we will obtain a binary string $(a_1, a_2,\cdots, a_n)$ with probabilities $(p_1, p_2, \cdots, p_n)$, where $a_i\in\{0, 1\}$, $p_i\in(0.5, 1]$ and $n$ indicates the number of qubits equal to $\beta q$. We select the $\lceil \log_2 n\rceil$ bits with probabilities closest to $0.5$ and generate $n$ candidate solutions by enumerating all possible combinations of these bits.
Furthermore, we perform coordinate restoration on these candidate solutions and calculate the  $\ell_2$ norm corresponding to each solution, finally select the vector with the smallest norm as the optimal solution.

\section{\label{sec:complexity_analysis}Analysis of qubit requirements for VQKZ }

In the previous section, we presented the framework of VQKZ and the construction method of the quantum SVP oracle. According to Algorithm \ref{alg:VQKZ}, the required number of qubits for solving SVP by using VQKZ is determined by the quantum SVP oracle. In this section, we conduct an detailed analysis of the qubit requirement for implementing the SVP oracle, thereby establishing an upper bound on the number of qubits required by the VQKZ.

For a lattice $\mathcal{L}(B)$ with $B=[\bm{b}_1,\bm{b}_2,\cdots,\bm{b}_r]$ and $\beta \geq 2$, considering its projected sublattice $\mathcal{L}(\hat{B}_{[j,k]})$ with  $\hat{B}_{[j,k]}=[\hat{\bm{b}}_j, \hat{\bm{b}}_{j+1},\cdots,\hat{\bm{b}}_k]$, $1\leq j\leq r-1$, $k=\min(j+\beta-1, r)$, we can choose a radius $R$ such that the shortest vector  $\bm{v}=\sum_{i=j}^kx_i\hat{\bm{b}}_i\in\mathcal{L}(\hat{B}_{[j,k]})$ in the sublattice satisfying
\begin{equation}\label{Eq:radius}
  \|v\| \leq R.
\end{equation}
By using the Gram-Schmidt orthogonalization~(GSO), we can transform the basis $\hat{B}_{[j,k]}$ into a orthogonal basis $[\hat{\bm{b}}^*_j, \hat{\bm{b}}^*_{j+1},\cdots,\hat{\bm{b}}^*_k]$ and rewrite $\bm{v}$ into
\begin{equation}\label{Eq:GSO_v}
  \bm{v} = \sum_{i=j}^k \left(x_i + \sum_{l=i}^k \eta_{i,l}\right)\hat{\bm{b}}^*_i = \sum_{i=j}^k \alpha_i\hat{\bm{b}}_i^*,
\end{equation}
where $\eta_{i,l}$ represent the GSO coefficients and $\alpha_i=x_i+\sum_{l=i}^k\eta_{i,l}$. Since $\|\bm{v}\|\leq R$, we can obtain $ \|x_k\hat{\bm{b}}^*_k\|\leq R$ and the relation with $\|x_{k-1}\hat{\bm{b}}^*_{k-1}\|$ as
\begin{equation}
  \|x_{k-1}\hat{\bm{b}}^*_{k-1}+x_k\hat{\bm{b}}^*_k\|^2\leq|x_{k-1}|^2\|\hat{\bm{b}}^*_{k-1}\|^2+|x_k|^2\|\hat{\bm{b}}^*_k\|\leq R^2.
\end{equation}
Hence, we can derive an upper bound for $x_{k-1}$ as
\begin{equation}
  |x_{k-1}|^2\leq \frac{R^2-|x_k|^2\|\hat{\bm{b}}^*_k\|^2}{\|\hat{\bm{b}}^*_{k-1}\|^2}\leq \frac{R^2}{\|\hat{\bm{b}}^*_{k-1}\|^2}.
\end{equation}
Similarly, for $j\leq i \leq k$, we can obtain the bound for each $x_i$ of $\bm{v}$ as
\begin{equation}
  |x_i|^2 \leq \frac{R^2}{\|\hat{\bm{b}}^*_i\|^2} \iff |x_i|\leq  \frac{R}{\|\hat{\bm{b}}^*_i\|}.
\end{equation}

To futher analyze the upper bound of $x_i$, we need to use the following theorem.
\begin{theorem}\label{Thm:1}
For a lattice basis $B=[\bm{b}_1,\bm{b}_2,\cdots,\bm{b}_r]$ and the projected sublattice basis $\hat{B}_{[j,k]}=[\hat{\bm{b}}_j, \hat{\bm{b}}_{j+1},\cdots,\hat{\bm{b}}_k]$ with a given $\beta\geq 2$, $1\leq j \leq r-1$ and $k=\min(j+\beta-1, r)$, their GSO forms satisfy
\begin{equation}
  \hat{\bm{b}}^*_i = \bm{b}^*_i,
\end{equation}
where $j\leq i \leq k$ and $\hat{\bm{b}}^*_i$, $\bm{b}^*_i$ represent to the GSO forms of $\bm{b}_i$ and $\hat{\bm{b}}_i$, respectively.
\end{theorem}
The proof is shown in Appendix~\ref{appendix:proof}. For the projected sublattice $\mathcal{L}(\hat{B}_{[j,k]})$, the radius $R$ is conventionally selected as $\|\bm{b}^*_j\|$. We can hence use the Theorem~\ref{Thm:1} and rewrite the upper bound for $x_i$ as
\begin{equation}\label{Eq:xi_bound}
  |x_i|\leq R_i = \frac{\|\bm{b}^*_j\|}{\|\bm{b}^*_i\|}.
\end{equation}

For the quantum SVP oracle, we use $q$ qubits to construct the operator $Q_i$ as Eq.~\ref{Eq:Q_i} whose spectrum is $[-2^{q-1}+1, 2^{q-1}]$. Combining with Eq.~\ref{Eq:xi_bound}, the number of qubits $q$ for constructing $Q_i$ should satisfy $2^{q-1} \geq R_i$, i.e., $q\geq \lfloor \log_2 R_i \rfloor +1 $. 
The total number of qubits required to solve SVP on $\mathcal{L}(\hat{B}_{[j,k]})$ can be obtained by summing the upper bound of all $x_i$ as following
\begin{align}
  n & = \sum_{i=j}^k ( \lfloor \log_2 R_i \rfloor +1) \\
    & \leq \beta + \lfloor \log_2 \Pi_{i=j}^k R_i \rfloor \label{Eq:n}
\end{align}
with $\beta = k-j+1$.

For $\Pi_{i=j}^k R_i$, since the $\beta$-BKZ reduced basis also satisfies the $\delta$-LLL reduced criteria, we can obtain $\delta\|\bm{b}_{i-1}^*\|^2\leq \|\bm{b}_i^* + \mu_{i,i-1}\bm{b}_{i-1}^*\|^2$ with $|\mu_{i,j}|\leq\frac{1}{2}$, and further derive the relation
\begin{equation}\label{Eq:bi_relation}
  \|\bm{b}^*_i\|^2\geq (\delta -\frac{1}{4})\|\bm{b}^*_{i-1}\|^2.
\end{equation} 
By using Eq.~\ref{Eq:bi_relation}, we can bound the $\Pi_{i=j}^k R_i$ as 
\begin{align}
  \Pi_{i=j}^k R_i & = \Pi_{i=j}^k  \frac{\|\bm{b}^*_j\|}{\|\bm{b}^*_i\|} \\
  & \leq \Pi_{i=j}^k ( \delta - \frac{1}{4})^{-\frac{1}{2}(i-j)} \\
  & = ( \delta - \frac{1}{4})^{-\frac{1}{2}\sum_{i=j}^k(i-j)}\\
  & = ( \delta - \frac{1}{4})^{-\frac{1}{2}[\frac{1}{2}\beta(\beta-1)]}\\
  & = ( \delta - \frac{1}{4})^{-\frac{1}{4}\beta(\beta-1)}.
\end{align}
Inserting this bound into Eq.~\ref{Eq:n}, we can finally derive the required number of qubits to construct the quantum SVP oracle as 
\begin{align}
  n &\leq \beta + \lfloor \log_2 ( \delta - \frac{1}{4})^{-\frac{1}{4}\beta(\beta-1)} \rfloor \\
  & = \beta - \lfloor\frac{1}{4}\beta(\beta-1)\log_2(\delta-\frac{1}{4}) \rfloor.
\end{align}
Specially, when $\delta = \frac{3}{4}$, the required number of qubits for VQKZ to solve SVP is equal to $n=\lfloor \frac{1}{4}(\beta^2+3\beta) \rfloor$.

\section{Effectiveness of VQKZ in solving typical SVP}

In Sec.~\ref{sec:VQKZ}, we construct the quantum SVP oracle build upon VQE and present the VQKZ to generate the $\beta$-BKZ reduced basis for solving SVP with $\lfloor \frac{1}{4}(\beta^2+3\beta) \rfloor$ qubits. In this section, we conduct simulation experiments to validate the efficacy of the SVP oracle, and further demonstrate that VQKZ outperforms two existing variational quantum SVP algorithms in both performance and the required number of qubit when solving typical SVP instances.

\section{Effectiveness of VQKZ in solving typical SVP}

In Sec.~\ref{sec:VQKZ}, we construct the quantum SVP oracle build upon VQE and present the VQKZ to generate the $\beta$-BKZ reduced basis for solving SVP with $\lfloor \frac{1}{4}(\beta^2+3\beta) \rfloor$ qubits. In this section, we conduct simulation experiments to validate the efficacy of the SVP oracle, and further demonstrate that VQKZ outperforms two existing variational quantum SVP algorithms in both performance and the required number of qubit when solving typical SVP instances.

\subsection{Effectiveness of quantum SVP oracle in generating  the first excited states}\label{sec:svp_oracle_implementation}

We first conducte experiments to demonstrate the effectiveness of the quantum SVP oracle. According to Sec.~\ref{quantum_svp_oracle}, given a lattice $\mathcal{L}(B)$ of rank $r$ and a block size $\beta \in [2, r-1]$, the quantum SVP oracle aims to generate the first excited state of the Hamiltonian $H$ corresponding to projected sublattices $\mathcal{L}(\hat{B}_{[j,k]})$ with $1\leq j\leq r-1$, $k=\min(j+\beta-1, r)$ . To demonstrate the ability of the quantum SVP oracle, we randomly generate $100$ lattices with $\beta=3, 4, 5$, respectively, and further utilize the SVP oracle to generate the first excited states of the Hamiltonian $H$ corresponding to these lattice.

\begin{figure*}[htp]
  \centering
  \subfigure[]
  {
      \centering
      \includegraphics[width=0.6\columnwidth]{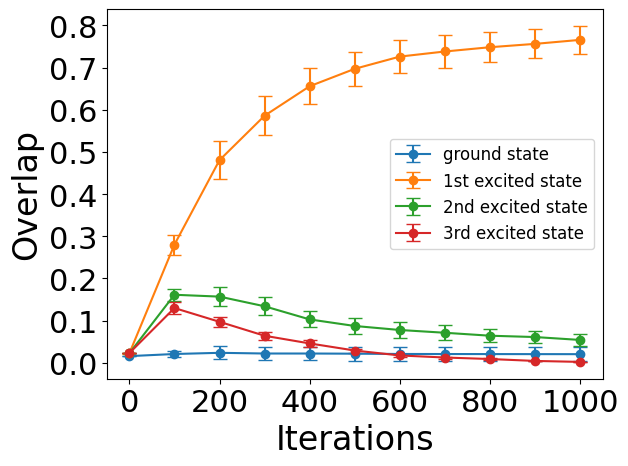}
      \label{oracle_d_3}
  }
  \subfigure[]
  {
      \centering
      \includegraphics[width=0.6\columnwidth]{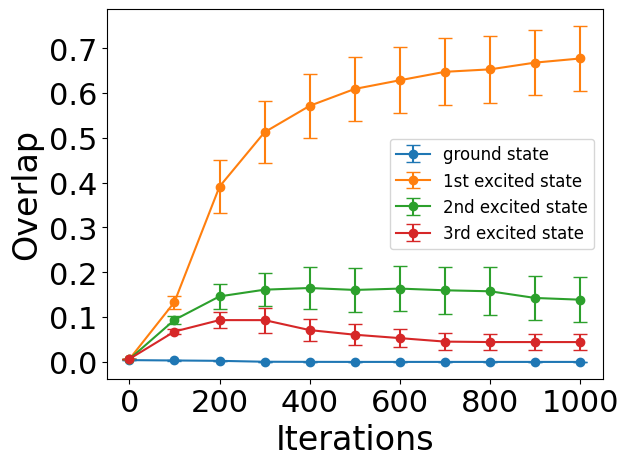}
      \label{oracle_d_4}
  }
  \subfigure[]
  {
      \centering
      \includegraphics[width=0.6\columnwidth]{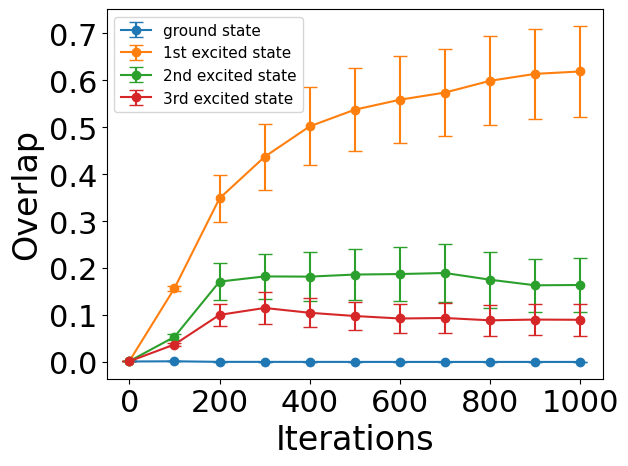}
     
      \label{oracle_d_5}
  }
  \caption{Simulation results for the quantum SVP oracle in VQKZ for generating the first excited states of the Hamiltonian corresponding to random lattices. \textbf{(a)} Results for lattice rank $\beta=3$. \textbf{(b)} Results for lattice rank $\beta=4$. \textbf{(c)} Results for lattice rank $\beta=5$.}
  \label{fig:svp_oracle}
\end{figure*}

For each Hamiltonian $H$, we firstly utilize the exact diagonalization method to obtain its  ground state and first three excited states, then compute the overlap between the quantum states generated by the SVP oracle and the eigenstates of $H$ with the following equation 
\begin{equation}
  dist(|\psi\rangle, |\phi\rangle):=|\langle \psi|\phi\rangle|^2,
\end{equation}
thereby quantifying the capability of the quantum SVP oracle  in producing the first excited state. Furthermore, we  construct each $Q_i$ in $H$ using two qubits according to Eq.~\ref{Eq:Q_i}. For $\beta=3, 4, 5$, we utilize the quantum circuits as describled in Fig.~\ref{oracle_ansatz} with $6$, $8$, $10$ layers, respectively. After $1000$ iterations, the quantum SVP oracle demonstrates high-probability occupation of the first three excited states in its output, with probabilities of $84.26\%$, $84.13\%$, and $75.04\%$ for $\beta=3, 4, 5$. These results demonstrate the oracle can efficiently generate short vectors in lattices. Specially, it can be obtained from Fig.~\ref{oracle_ansatz} that the quantum SVP oracle can generate the first excited state, i.e. the shortest vector,  with maximal probabilities of $76.59\%$, $70.23\%$, $61.89\%$ for $\beta=3, 4, 5$. This result demonstrates that the shortest vector on the projected sublattice can be effectively obtained through a few measurement operations.

\subsection{Comparison with two variational quantum SVP algorithms}

\subsubsection{Theoretical comparison}
\begin{figure}[htp]
  \setlength{\belowcaptionskip}{-2mm}
  \centering
  
  \includegraphics[width=0.8\linewidth]{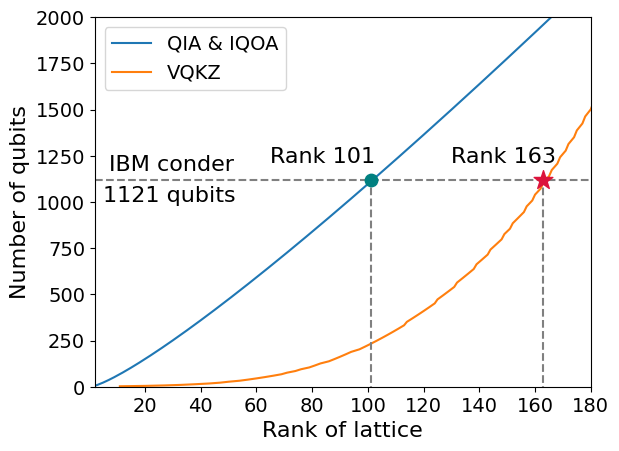}
  
  \caption{A theoretical comparison of the qubit requirements for VQKZ, QIA, and IQOA in relation to lattice rank.}
  \label{theoretical_comparison}
\end{figure}

In addition to validating the capability of quantum SVP oracle to produce the first excited states of Hamiltonian $H$, we conduct comparative analyses between VQKZ and two established variational quantum SVP methods, the quantum Ising algorithm (QIA)~\cite{joseph2021two} and the iterative quantum optimization algorithm (IQOA)~\cite{zhu2022iterative}, to demonstrate the advantages of VQKZ in solution quality for typical SVPs and resource efficiency through reduced qubit requirements. QIA and IQOA are two variational quantum algorithms designed to solve the SVP by mapping it onto a fully connected quantum Ising model. Specifically, for a lattice basis $B = [\bm{b}_1, \cdots, \bm{b}_r]$, they generate the Hamiltonian as 
\begin{equation}
  H = \sum_{i=1}^r\sum_{j=1}^r \bm{b}_i \cdot \bm{b}_j^T Q_iQ_j,
\end{equation}
where $Q_i$ is defined as Eq.~\ref{Eq:Q_i}. According to Ref.~\cite{joseph2021two}, when the lattice satisfies the Hermite Normal Form (HNF), the number of qubits required for QIA and IQOA can be given as $1.5r\log_2r+r+\log_2D$ where $D$ represents the covolume of lattice defining as $D=\text{det}(\mathcal{L})$. Since the value of $D$ depends on the specific lattice basis, we ignore it and take $n\approx1.5r\log_2r+r$ as the number of qubits for QIA and IQOA. This allows us to derive the functional relationship between the number of qubits and the rank of the lattice in the SVP, as shown by the blue line in Fig.~\ref{theoretical_comparison}.

For VQKZ, according to Ref.~\cite{li2025complete}, the Euclidean norm of the first vector $\bm{b}_1$ in the $\beta$-BKZ reduced basis satisfies
\begin{equation}
  \|\bm{b}_1\|\leq \gamma_\beta^{\frac{r-1}{2(\beta-1)}+\frac{\beta(\beta-2)}{2r(\beta-1)}}|\text{det}(\mathcal{L})|^{1/r},
\end{equation}
where $\gamma_\beta$ denotes as the Hermite constant for the lattice, that can be bounded by $\gamma_\beta < \frac{r}{8}+\frac{6}{5}$~\cite{wen2018kz}. Furthermore, in lattice analysis, the Gaussian heuristic is a widely used estimation for the length of the shortest nonzero vector in a lattice, which can be expressed as 
\begin{equation}
  \text{GH}(\mathcal{L})\approx \sqrt{\frac{r}{2\pi e}}|\text{det}(\mathcal{L})|^{1/r}.
\end{equation}
Let the upper bound of the norm of $\bm{b}_1$ be smaller than the Gaussian heuristic. Then, for a given $\beta$, the rank of the lattice in the SVP that VQKZ can solve should satisfy
\begin{equation}
  \left(\frac{r}{8}+\frac{6}{5}\right)^{\frac{r-1}{2(\beta-1)}+\frac{\beta(\beta-2)}{2r(\beta-1)}}-\sqrt{\frac{r}{2\pi e}}\leq 0.
\end{equation}
For the chosen $\beta$, the number of qubits required by VQKZ is $\lfloor \frac{1}{4}(\beta^2+3\beta)  \rfloor$. Therefore, as the rank of the lattice increases, the number of qubits required by VQKZ can be determined, resulting in the orange line shown in Fig.~\ref{theoretical_comparison}. It can be obtained that as the lattice rank increases, the number of qubits required by VQKZ is significantly lower than that required by QIA and IQOA. This demonstrates that VQKZ can effectively reduce the scale requirements of near-term quantum devices for solving the SVP problem. Furthermore, considering the current quantum computing hardware, IBM Condor provides 1121 qubits~\cite{condor}. With this device, QIA and IQOA can theoretically solve SVP problems on lattices with a rank of 101, while VQKZ can solve SVP problems on lattices with a rank of 163, representing a $61.39\%$ increase in problem size.

It is worth mentioning that the comparison shown in Fig.~\ref{theoretical_comparison} only demonstrates the minimum number of qubits required in theory but cannot determine the complexity of QIA, IQOA, and VQKZ in solving SVP. Since all three algorithms are variational quantum algorithms, their complexity is jointly determined by the number of qubits, the circuit depth, and the number of iterations required to generate the first excited state. For the circuit depth, the number of quantum gates required by variational quantum algorithms for circuit depth typically scales polynomially with the number of qubits. This implies that solving the SVP using IBM Condor still requires thousands or even tens of thousands of quantum gates. With the current quantum gate error rate of approximately $10^{-3}$, the probability of the quantum computer yielding a correct solution remains extremely low, posing significant challenges for executing the optimization process. Therefore, although the number of qubits in current quantum computing devices can theoretically meet the requirements for solving large-scale SVP instances, there still remains a significant gap between this theoretical capability and the practical realization of solving SVP problems using real quantum hardware. Further theoretical analysis of circuit depth and iteration will be investigated in future research. In the next subsection, we compare the performance of the three algorithms in solving SVP instances through numerical simulations.

\subsubsection{Simulation comparison}

In addition to theoretical comparisons, we conducted simulation experiments based on SVP instances of random lattices and $q$-ary lattices to compare the performance and the number of qubits required by VQKZ, QIA, and IQOA. To evaluate algorithmic performance, we use the output vector of classical BKZ algorithm as the baseline, computing relative errors $\epsilon$ between the $\ell_2$-norm of the shortest vectors obtained by each quantum algorithm and those generated by the classical BKZ method. Denoting the $\ell_2$-norm of the shortest vector generated by the classical BKZ algorithm as $\lambda$ and the $\ell_2$-norm of the shortest vector obtained from each quantum algorithm as $\lambda_1(\cdot)$, we define the relative error $\epsilon$ as

\begin{equation}
  \epsilon := \left|\frac{\lambda-\lambda_1(\cdot)}{\lambda}\right|\times 100\%.
\end{equation}

\begin{figure}[htp]
  \centering
  \subfigure[]
  {
      \centering
      \includegraphics[width=0.8\columnwidth]{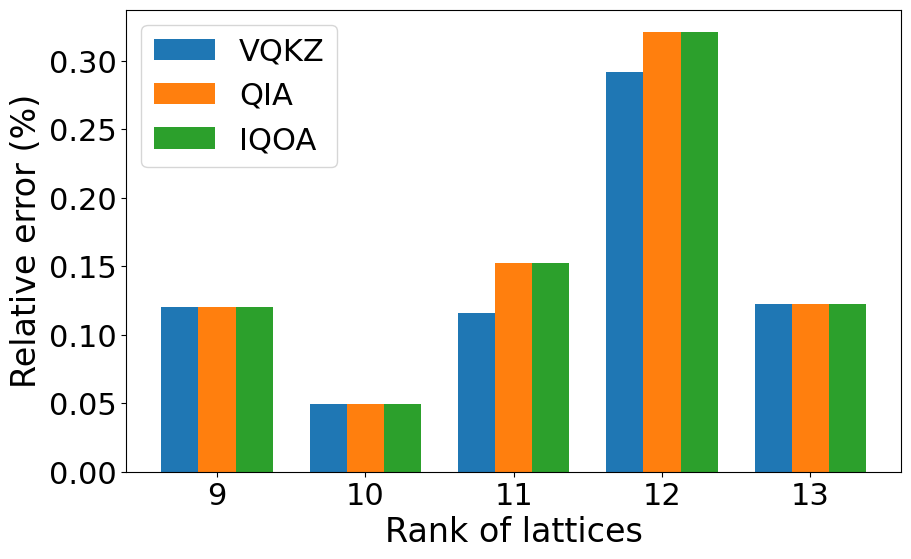}
      \label{fig:comparison_random}
  }
  \subfigure[]
  {
      \centering
      \includegraphics[width=0.8\columnwidth]{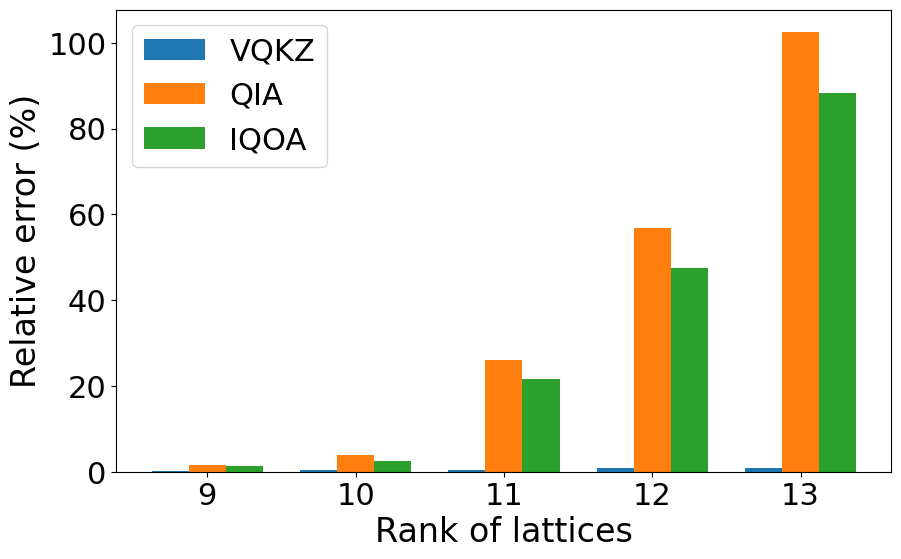}
      \label{fig:comparison_q_ary}
  }
  
  \caption{Comparison of the average relative errors of VQKZ, QIA, and IQOA in solving typical SVP problems. \textbf{(a)} Results for random lattices with rank ranging from 9 to 13. \textbf{(b)} Results for the $q$-ary lattices with rank ranging from 9 to 13. }
  \label{fig:comparison}
\end{figure}

Using the relative error metric as the baseline, we first evaluate the three quantum algorithms on SVP instances constructed from random lattices. For random lattices, we construct $\mathcal{L}(B)$ from random integer matrices of rank ranging from $9$ to $13$, generating $100$ lattices per rank. For VQKZ, we adopt a block size $\beta=3$ for all lattices and use $2$ qubits to construct $Q_i$ as shown in Eq.~\ref{Eq:Q_i}. Meanwhile, we use $2$ qubits to construct the coordinate search space of each lattice basis for QIA and IQOA as introduced in Ref.~\cite{joseph2021two} and Ref.~\cite{zhu2022iterative}. Results in Fig.~\ref{fig:comparison_random} show the averaged relative errors across $100$ lattices per rank, demonstrating that VQKZ achieves comparable $\ell_2$-norm precision to QIA and IQOA on random lattices. Specially, as demonstrated in Table~\ref{num_qubits}, VQKZ solves these SVP instances using only $6$ qubits, representing a significant reduction in qubit count compared to QIA and IQOA. 

\begin{table}[htp]
  \centering
  \begin{tabular}{c|c|c|c|c|c}
  \hline
  \hline
  \diagbox{Model}{\# Qubits}{Rank} & $9$ & $10$ & $11$ & $12$ & $13$ \\
  \hline 
  VQKZ & $6$ & $6$ & $6$ & $6$ & $6$ \\
  \hline
  QIA & $18$ & $20$ & $22$ & $24$ & $26$\\
  \hline
  IQOA & $18$ & $20$ & $22$ & $24$ & $26$\\
  \hline
  \hline
  \end{tabular}
  \caption{The required number of qubits for three quantum SVP algorithms for solving the SVP instances based on random lattices and the $q$-ary lattices. }
  \label{num_qubits}
\end{table}

Besides random lattices, we further implement the comparison among three quantum SVP algorithms based on the $q$-ary lattices. The $q$-ary lattice is widely used in post-quantum cryptography and can be describled as   
\begin{equation}
  \mathcal{L}(B):= \left(\begin{matrix}
    I_{d-k} & A \\
    \bm{0} & q\cdot I_k
  \end{matrix}\right),  
\end{equation}
where $A\in\mathbb{Z}_q^{(d-k)\times k}$ and $q\in\mathbb{N}^+$. Similar with the comparison on the random lattices, we randomly generate $100$ $q$-ary lattices for each rank, which ranges from $9$ to $13$. It can been seen from Fig.~\ref{fig:comparison_q_ary} and Table~\ref{num_qubits} that the VQKZ achieves much lower relative error than QIA and IQOA with fewer number of qubits. Specially, for the SVP based on the $q$-ary lattices with rank $13$, the VQKZ achieves the average relative error $0.86\%$ with only $6$ qubits, significant lower than $88.24\%$ achieved by IQOA and $102.60\%$ achieved by QIA, which use $26$ qubits.These results demonstrate that the VQKZ can effectively solve the typical SVP  and is more likely to be implemented on near-term quantum devices.

\section{Conclusion}
In this work, we propose a variational quantum Korkin-Zolotarev (VQKZ) algorithm for efficiently solving the SVP on near-term quantum devices. Unlike existing variational quantum methods whose required number of qubits increases with the SVP problem size, the VQKZ algorithm reduces the qubit requirements by transforming the original problem into a series of subproblems on projected sublattices of smaller rank $\beta$. Specifically, the algorithm leverages a VQE-based oracle to construct a reduced basis satisfying the $\beta$-BKZ reduced criteria, with an upper bound of only $\lfloor\frac{1}{4}(\beta^2 + 3\beta)\rfloor$ qubits.

Numerical simulations demonstrate that the VQKZ algorithm not only reduces qubit requirements but also outperforms existing variational quantum approaches in finding shorter vectors for typical SVP instances, including those based on random lattices and q-ary lattices. The proposed SVP ansatz and classical post-processing strategy further enhance the performance of the SVP oracle, enabling the efficient generation of the first excited states of the Hamiltonian corresponding to projected sublattices. Moreover, the simulations reveal that VQKZ achieves comparable or superior solution precision to other methods while requiring significantly fewer qubits, making it more practical for near-term quantum devices.

\section*{Acknowledgments}

We acknowledge financial support from the National Key Research and Development Program of China (Grant No. 2020YFA0309704), the National Natural Science Foundation of China (Grants No.62471446, No. 62171418, No. 62201530, No.U24B20135, and No. 62301517), the Sichuan Science and Technology Program (Grants No.2024JDDQ0008, No. 2023ZYD0131, No. 2023JDRC0017, No. 2022ZDZX0009, No. 2024NSFSC0470, No. 2024NSFSC0454, No.2024ZYD0008, and No.2025ZNSFSC1473), the National Key Laboratory of Security Communication Foundation(6142103042301, 6142103042406, 6142103042402), Stability Program of National Key Laboratory of Security Communication(WD202413, WD202414).

\bibliographystyle{unsrt}
\bibliography{ref}

\clearpage

\appendix

\section{Proof of Theorem 1}\label{appendix:proof}
 
For a given block size $\beta \geq 2$ and a lattice $\mathcal{L}(B)$ with the basis $(B)=[\bm{b}_1, \bm{b}_2, \cdots, \bm{b}_r]$, the basis of its projected sublattice $\hat{B}_{[j, k]}=[\hat{\bm{b}}_j, \hat{\bm{b}}_{j+1}, \cdots, \hat{\bm{b}}_{k}]$ with $1\leq j \leq r-1$ and $k=\min(j+\beta-1, r)$ can be described as
\begin{gather*}
  \hat{\bm{b}}_j = \pi_j(\bm{b}_j) = \bm{b}_j^*, \\
  \hat{\bm{b}}_{j+1} = \pi_j(\bm{b}_{j+1}) = \mu_{j+1, j}\bm{b}_j^* + \bm{b}_{j+1}^*,\\
  \hat{\bm{b}}_{j+2} = \pi_j(\bm{b}_{j+2}) = \mu_{j+2, j}\bm{b}_j^* + \mu_{j+2, j+1}\bm{b}_{j+1}^* + \bm{b}_{j+2}^*,\\
  \vdots\\
  \hat{\bm{b}}_k = \pi_j(\bm{b}_k) = \sum_{i=j}^{k-1}\mu_{k, i} \bm{b}_i^* + \bm{b}_k^*,
\end{gather*}
where $\bm{b}_i^*$ represents the Gram-Schmidt orthogonalized vectors of the lattice basis and $\mu_{i,j}$ indicates the Gram-Schmidt orthogonalization coefficients. Labeling the Gram-Schmidt orthogonalized vectors of $\hat{B}_{[j, k]}$ as $\{\hat{\bm{b}}^*_i\}_{i=j}^k$ with the coefficients $\{\eta_{i, l}\}$, $j\leq l< i\leq k$, we can obtain the following equation for $\hat{\bm{b}}^*_j$ and $\bm{b}_j^*$ as 
\begin{equation}
  \hat{\bm{b}}^*_j = \hat{\bm{b}}_j = \pi_j(\bm{b}_j) = \bm{b}_j^*,
\end{equation}
where $\hat{\bm{b}}^*_j = \hat{\bm{b}}_j$ can be obtained by using the definition of the Gram-Schmidt orthogonalization. 

For all $i$ with $j\leq i \leq k$, without loss of generality, we assume that $\hat{\bm{b}}_l^* = \bm{b}_l^*$ holds for $j\leq l \leq i-1$. Meanwhile, for $\hat{\bm{b}}_i^*$, we can obtain that
\begin{align}
  \hat{\bm{b}}_i^* & = \hat{\bm{b}}_i - \sum_{l=j}^{i-1} \eta_{i, l} \hat{\bm{b}}_l^*\\
  &= \pi_j(\bm{b}_i) - \sum_{l=j}^{i-1} \eta_{i, l} \hat{\bm{b}}_l^*\\
  &= \sum_{m=j}^{i-1} \mu_{i, m}\bm{b}_m^* + \bm{b}_i^* - \sum_{l=j}^{i-1} \eta_{i, l} \hat{\bm{b}}_l^*.\label{Eq:bi}
\end{align}
For $\eta_{i, l}$, we have
\begin{align}
  \eta_{i, l} &= \frac{\hat{\bm{b}}_i\cdot \hat{\bm{b}}_l^*}{ \hat{\bm{b}}_l^* \cdot  \hat{\bm{b}}_l^*}\\
  & = \frac{(\sum_{m=j}^{i-1}\mu_{i, m}\bm{b}^*_m + \bm{b}_i^*)\cdot \bm{b}_l^*}{\bm{b}_l^* \cdot \bm{b}_l^*}\\
  &= \frac{\sum_{m=j}^{i-1}\mu_{i, m}\bm{b}^*_m \cdot \hat{\bm{b}}_l^*}{\hat{\bm{b}}_l^* \cdot \hat{\bm{b}}_l^*} \\
  & = \mu_{i,l},
\end{align}
where, in the third line, we use the orthogonal property of the Gram-Schmidt basis. Inserting $\eta_{i, l} = \mu_{i,l}$ into Eq.~\ref{Eq:bi}, we can finally obtain
\begin{align}
  \hat{\bm{b}}_i^* & = \sum_{m=j}^{i-1} \mu_{i, m}\bm{b}_m^* + \bm{b}_i^* - \sum_{l=j}^{i-1} \mu_{i, l} \hat{\bm{b}}_l^*\\
  &= \bm{b}_i^*,
\end{align}
which completes the proof. 

\section{Comparison between SVP ansatz and hardware-efficient ansatz}\label{appendix:comparison}

\begin{figure}[H]
  \setlength{\belowcaptionskip}{-2mm}
  \centering
  
  \includegraphics[width=1\linewidth]{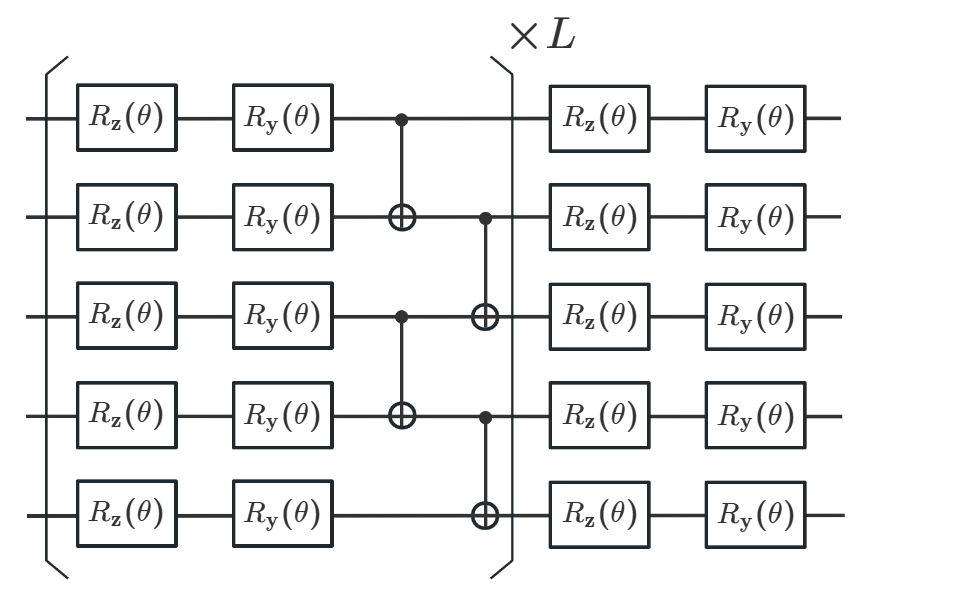}
  
  \caption{An hardware-efficient ansatz with $5$ qubits and $L$ layers.}
  \label{fig:HEA}
\end{figure}

In this section, we compare the performance of the SVP ansatz and the hardware-efficient~(HE) ansatz, as shown in Fig.~\ref{fig:HEA}, when applied to the quantum SVP oracle in generating the first excited state of the Hamiltonian $H$ shown in Eq.~\ref{Eq:H}, and further demonstrate the advantages of the SVP ansatz in reducing the number of quantum gates and improving effectiveness. To perform the comparison, we adopt the similar experimental setup as described in Sec.\ref{sec:svp_oracle_implementation}. We first generated $100$ random matrices with ranks $3$, $4$, and $5$, respectively. For each lattice $\mathcal{L}(B)$, we construct the corresponding Hamiltonian $H$, and employ quantum SVP oracles based on the SVP ansatz and the HE ansatz. After $1000$ iterations, we calculate the overlaps between the quantum states produced by the oracles and the ground state, as well as the first two excited states of $H$, and then compare the performance of the quantum SVP oracle based on the two ansatze.

\begin{figure}[htp]
  \centering
  \subfigure[]
  {
      \centering
      \includegraphics[width=\columnwidth]{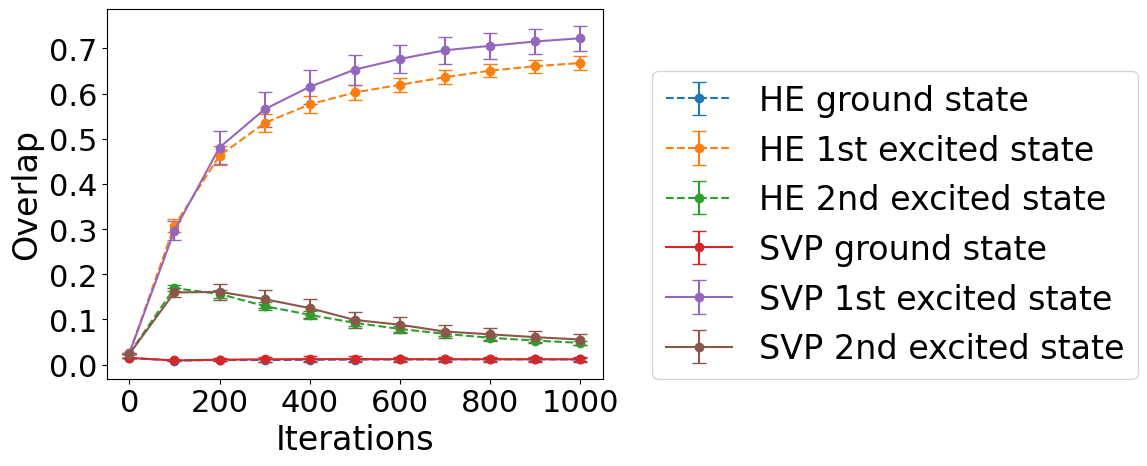}
      \label{comparison_d_3}
  }
  \subfigure[]
  {
      \centering
      \includegraphics[width=\columnwidth]{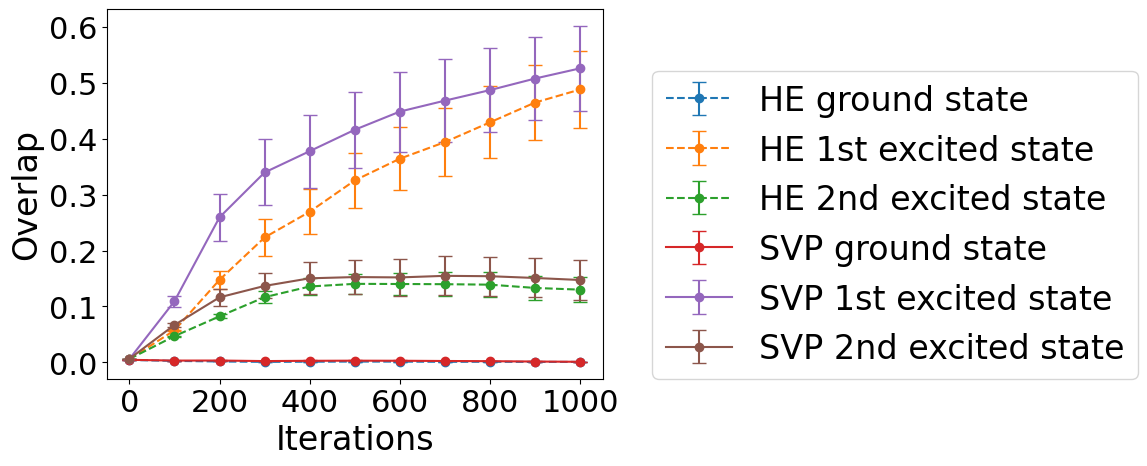}
      \label{comparison_d_4}
  }
  \subfigure[]
  {
      \centering
      \includegraphics[width=\columnwidth]{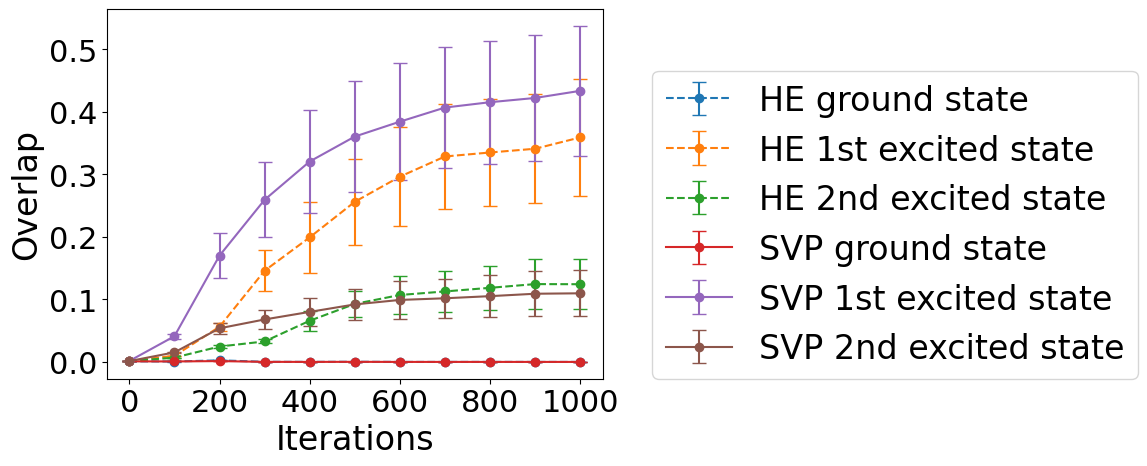}
     
      \label{comparison_d_5}
  }

  \caption{Comparison of SVP ansatz and HE ansatz in generating the first-excited states of Hamiltonian corresponding to random lattices.\textbf{(a)} Results for lattice with $rank = 3$. \textbf{(b)} Results for lattice with $rank = 4$. \textbf{(c)} Results for lattice with $rank = 5$.}
  \label{fig:comparison_hea}
\end{figure}

For the two circuit ansatz, we chose a $10$-layer HE ansatz for each task and then chose an SVP ansatz with a similar number of parameters to the HE ansatz. The detailed configurations of the two ansatz are shown in Table~\ref{Tab:comparison}. The experimental results are shown in Fig.\ref{fig:comparison_hea}. It can be observed that, across three different lattice sizes, the SVP ansatz demonstrates a higher probability of effectively generating the first excited state compared to the HE ansatz. Moreover, for lattices with $r=3$, $4$, and $5$, the number of quantum gates required by the SVP ansatz is reduced by $22.53\%$, $25.20\%$, and $12.90\%$, respectively, compared to the HE ansatz. These results indicate that the SVP ansatz can effectively reduce circuit complexity when generating the first excited state, making it more suitable for implementation on near-term quantum computing devices.

\begin{table}[htp]
  \centering
  \setlength{\belowcaptionskip}{-.3cm}
  \begin{tabular}{c|c|c|c|c}
  \hline
  \hline
  Rank & Ansatz & \# Layers & \# Parameters & \# Gates \\
  \hline
  \multirow{2}*{$r=3$} & SVP & $5$ & $135$ & $141$  \\
  \cline{2-5}
  ~ & HE & $10$ & $132$ & $182$ \\
  \hline
  \multirow{2}*{$r=4$} & SVP & $4$ & $176$ & $184$  \\
  \cline{2-5}
  ~ & HE & $10$ & $176$ & $246$ \\
  \hline
  \multirow{2}*{$r=5$} & SVP & $4$ & $260$ & $270$  \\
  \cline{2-5}
  ~ & HE & $10$ & $220$ & $310$ \\
  \hline
  \hline
  \end{tabular}
  \caption{Circuit configurations for the SVP ansatz and HE ansatz in comparison experiments.}
  \label{Tab:comparison}
\end{table}

\end{document}